\documentclass[aps,prl,showpacs,twocolumn,floatfix]{revtex4}

\usepackage{amsmath}
\usepackage{bm}
\usepackage{mathrsfs}
\usepackage{graphicx}
\usepackage{subfigure}

\begin{document}

\title{Dynamics of spin $\tfrac{1}{2}$ quantum plasmas}
\author{Mattias Marklund\footnote{Electronic address: \texttt{mattias.marklund@physics.umu.se}}$^\ddag$ 
  and Gert Brodin\footnote{Electronic address: \texttt{gert.brodin@physics.umu.se}}}
\altaffiliation[Also at: ]{Centre for Fundamental Physics, Rutherford Appleton Laboratory, Chilton, 
  Didcot, Oxon OX11 OQX, U.K.}
\affiliation{Department of Physics, Ume{\aa} University, SE--901 87 Ume{\aa}, Sweden}

\received{November 8, 2006}
\date{\today}
\revised{December 1, 2006}

\begin{abstract}
  The fully nonlinear governing equations for spin $\tfrac{1}{2}$ quantum plasmas are presented. 
  Starting from the
  Pauli equation, the relevant plasma equations are derived, and it is shown that nontrivial quantum 
  spin couplings arise, enabling studies of the combined collective and spin dynamics. 
  The linear response of the quantum plasma in an electron--ion system is obtained and analyzed. 
  Applications of the theory to solid state and astrophysical systems as well as dusty plasmas are 
  pointed out.  
\end{abstract}

\maketitle

There is currently a great deal of interest in investigating collective
plasma modes \cite
{haas-etal1,anderson-etal,haas-etal2,haas,garcia-etal,marklund,Shukla-Eliasson,marklund-shukla}
in quantum plasmas, as such plasmas could be of relevance in nano-scale
electro-mechanical systems \cite{Markowich-etal,Calvayrac-etal,Stenflo-etal},
in microplasmas and dense laser-plasmas \cite{Becker-etal}, and 
laser interactions with atomic systems \cite{exp1,exp2}. For
example, Refs.\ \cite{haas-etal1} and \cite{haas-etal2,haas,garcia-etal}
used quantum transport models in order to derive modified dispersion
relations for Langmuir and ion-acoustic waves, while Shukla \& Stenflo \cite
{shukla-stenflo} ionvestigated drift modes in nonuniform quantum
magnetoplasmas. Moreover, it is known that cold quantum plasmas can support
new dust modes \cite{shukla-mamun,shukla}. In Ref.\ \cite{Shukla-Eliasson}
it was shown that electron quantum plasmas could
support highly stable dark solitons and vortices. Further examples of
quantum plasmas and the range of validity of their descriptions has been
discussed recently in Ref.\ \cite{manfredi}. The above studies of
quantum plasmas have used models based on the Schr\"odinger description of
the electron. It is expected that new and possible important effects could
appear as further quantum effects are incorporated in models describing
the quantum plasma particles. The coupling of spin to classical motion has 
attracted interest in the literature (see, e.g., \cite{halperin-hohenberg,blum,balatsky,rathe-etal,hu-keitel,%
arvieu-etal,aldana-roso,walser-keitel,qian-vignale,walser-etal,roman-etal,liboff,fuchs-etal}). Much work has 
been done concerning
single particle spin effects in external field configurations, such as intense laser fields \cite{rathe-etal,hu-keitel,arvieu-etal,aldana-roso,walser-keitel,walser-etal}, and 
the possible experimental signatures thereof. However, there have also been interest in 
excitations of collective modes in spin systems, such as spin waves, in a wide scientific community. 
For example, in Refs.\ \cite{halperin-hohenberg,blum,balatsky} hydrodynamical 
models including spin was presented, and further theory concerning spin, angular momentum, and the 
forces related to spin was discussed in Refs.\ \cite{roman-etal} and \cite{liboff}. Moreover, spin waves in spinor Bose condensates has recently been discussed in, e.g., Ref.\ \cite{fuchs-etal}. 
The treatment of charged particles and plasmas
using quantum theory has received attention in astrophysical settings, especially in
strongly magnetized environments \cite{melrose,harding-lai}. For example, effects of quantum field theory
on the linear response of an electron gas has been analyzed \cite{melrose-weise}, results concerning 
the spin-dependence of cyclotron decay on strong magnetic fields has been 
presented \cite{baring-etal}, and the propagation of quantum electrodynamical waves
in strongly magnetized plasmas has been considered \cite{brodin-etal}.

In this Letter we present for the first time the fully nonlinear governing equations for spin $\tfrac{1}{2}$ quantum
electron plasmas. Starting from the Pauli
equation describing the nonrelativistic electron, we show that the electron--ion
plasma equations are subject to spin related terms. These terms
give rise to a multitude of collective effects of which some are investigated in detail. 
Applications of the 
governing equations are discussed, and it is shown that under certain circumstances the collective spin
effects can dominate the plasma dynamics. 

We will assume that the electron wave function can be written in the product form 
$\Psi = \varPsi_{(1)}\varPsi_{(2)}\ldots\varPsi_{(N)}$, where $N$ is the number of 
particle states. Thus we will here neglect the effects of entanglement and focus on the collective properties of
the the quantum electron plasma. Then   
the non-relativistic evolution of spin $\tfrac{1}{2}$ particles, as described by the 
two-component spinor $\varPsi_{(\alpha)}$, is given by (see, e.g.\ \cite{holland})
\begin{equation}\label{eq:pauli}
  i\hbar\frac{\partial\varPsi_{(\alpha)}}{\partial t} = \left[ 
    -\frac{\hbar^2}{2m_e}\left( \bm{\nabla} + \frac{ie}{\hbar c}\bm{A} \right)^2
    + \mu_B\bm{B}\cdot\bm{\sigma} - e\phi
  \right] \varPsi_{(\alpha)} 
\end{equation}
where $\alpha$ numbers the particle states, $m_e$ is the particle mass, $\bm{A}$ is the vector potential, $e$ is the magnitude of the electron charge, $\mu_B =-e\hbar/2m_ec$ is the electron magnetic moment, 
$\phi$ is the electrostatic potential, and 
$\bm{\sigma} = (\sigma_1, \sigma_2, \sigma_3)$ are 
the Pauli spin matrices, represented by 
\begin{equation}
  \sigma_1 = \left( 
  \begin{array}{cc}
    0 & 1 \\
    1 & 0
  \end{array}
  \right) , \,
  \sigma_2 = \left(
  \begin{array}{cc}
    0 & -i \\
    i & 0
  \end{array} 
  \right) , \, \text{ and }\,
  \sigma_3 = \left(
  \begin{array}{cc}
    1 & 0 \\
    0 & -1
  \end{array} 
  \right) .
\end{equation}  

By introducing the decomposition of the spinors according to
$
  \varPsi_{(\alpha)} = \sqrt{ n_{(\alpha)}}\,\exp(iS_{(\alpha)}/\hbar)\varphi_{(\alpha)} , 
$ 
 we may derive a set of $N$ coupled fluid equations \cite{holland} for the 
densities $n_{(\alpha)}$, the velocities $\bm{v}_{(\alpha)} = (1/m_e)\left( \bm{\nabla}S_{(\alpha)} 
    - i\hbar\varphi^{\dag}\bm{\nabla}\varphi \right) + (e/m_ec)\bm{A}$,  
and the spin vectors $\bm{s}_{(\alpha)} = (\hbar/2)\varphi_{(\alpha)}^{\dag}\bm{\sigma}\varphi_{(\alpha)}$ (where 
$\varphi_{(\alpha)}$ is
the 2-spinor through which the spin $\tfrac{1}{2}$ properties are 
mediated).

Next we define
the total particle density for the species with charge $q$ according to 
$
  n_e = \sum_{{(\alpha)} = 1}^Np_{\alpha}n_{(\alpha)} ,
$ 
where $p_{\alpha}$ is the probability related to the wave function $\varPsi_{(\alpha)}$. 
Using the ensemble average $\langle f\rangle = \sum_{\alpha}p_{\alpha}(n_{(\alpha)}/n)f$ for
any tensorial quantity $f$, we define the total electron fluid velocity for charges  $\bm{V}_e = \langle\bm{v}_{(\alpha)}\rangle$
and the total electron spin density $\bm{S} = \langle\bm{s}_{(\alpha)}\rangle $.
From these definitions we can define the microscopic
velocity in the electron fluid rest frame according to $\bm{w}_{(\alpha)} = \bm{v}_{(\alpha)} - \bm{V}_e$, satisfying
$\langle\bm{w}_{(\alpha)}\rangle = 0$, and the microscopic spin density $\bm{\mathcal{S}}_{(\alpha)} 
= \bm{s}_{(\alpha)} - \bm{S}$, such that $\langle\bm{\mathcal{S}}_{(\alpha)}\rangle = 0$.

We then obtain the conservation equations
\begin{equation}\label{eq:density}
  \partial_t n_e  + \bm{\nabla}\cdot(n_e\bm{V}_e) = 0 ,
\end{equation}
\begin{eqnarray}
&&
  mn_e\left( \partial_t + \bm{V}_e\cdot\bm{\nabla}\right)\bm{V}_e 
    = -en_e\left( \bm{E} + \bm{V}_e\times\bm{B} \right) \nonumber \\
&&\qquad
    - \bm{\nabla}\cdot\bm{\mathsf{\Pi}}_e - \bm{\nabla}P_e + \bm{\mathcal{C}}_{ei} + \bm{F}_Q
    \label{eq:mom}
\end{eqnarray}
and 
\begin{eqnarray}
  n_e\left( \partial_t + \bm{V}_e\cdot\bm{\nabla} \right)\bm{S}
  = \frac{2\mu_B n_e}{\hbar}\bm{B}\times\bm{S} - \bm{\nabla}\cdot{\bm{\mathsf{K}}}_e
  + \bm{\Omega}_S
\label{eq:spin-total}
\end{eqnarray}
respectively. 
Here we have added the electron--ion collisions $\bm{\mathcal{C}}_{ei}$, 
denoted the total quantum force density by
\begin{eqnarray}
&&\!\!\!\!\!\!\!   
   \bm{F}_Q =  - n_e\langle\bm{\nabla}Q_{(\alpha)}\rangle 
    - \frac{2\mu_B n_e}{\hbar}(\bm{\nabla}\otimes\bm{B})\cdot\bm{S} 
  \nonumber \\ &&\!\!\!\!\!\!\! \quad
    - \frac{1}{m_e}\bm{\nabla}\cdot\left( 
        n_e \bm{\mathsf{\Sigma}}\,
      \right)
     - \frac{1}{m_e}\bm{\nabla}\cdot\big( 
        n_e \widetilde{\bm{\mathsf{\Sigma}}}\,
      \big) 
  \nonumber \\ &&\!\!\!\!\!\!\! \quad
    - \frac{2}{m_e}\bm{\nabla}\cdot\left\{ 
        n_e\,\mathrm{Sym}\!\left[(\bm{\nabla}S_a)\otimes\langle (\bm{\nabla}\mathcal{S}^{a}_{(\alpha)})\rangle
      \right]\right\}  , 
\end{eqnarray}
where $\mathrm{Sym}$ denotes the symmetric part of the tensor, 
and defined the nonlinear spin fluid contribution by
\begin{eqnarray}
&&
  \bm{\Omega}_S =
    \frac{1}{m_e}\bm{S}\times[\partial_a(n_e\partial^a\bm{S})]
   + \frac{1}{m_e}\bm{S}\times[\partial_a(n_e\langle\partial^a\bm{\mathcal{S}}_{(\alpha)}\rangle)]
  \nonumber \\ && \qquad 
   + \frac{n_e}{m_e}\left\langle\frac{\bm{\mathcal{S}}_{(\alpha)}}{n_{(\alpha)}}\times[\partial_a(n_{(\alpha)}\right\rangle\partial^a\bm{{S}})]
    \nonumber \\ && \qquad 
   + \frac{n_e}{m_e}\left\langle\frac{\bm{\mathcal{S}}_{(\alpha)}}{n_{(\alpha)}}\times[\partial_a(n_{(\alpha)}\partial^a\bm{\mathcal{S}}_{(\alpha)})] \right\rangle ,
   \label{omega}
\end{eqnarray} 
where
$\bm{\mathsf{\Pi}}_e = m_en_e[\langle\bm{w}_{(\alpha)}\otimes\bm{w}_{(\alpha)}\rangle 
- \bm{\mathsf{I}}\langle w_{(\alpha)}^2\rangle /3 ]$ is the
trace-free anisotropic pressure tensor ($\bm{\mathsf{I}}$ is the unit tensor), 
$P_e = m_en_e\langle w_{(\alpha)}^2\rangle$ is the isotropic 
scalar pressure,  $\bm{\mathsf{\Sigma}} = (\bm{\nabla}S_a)\otimes(\bm{\nabla}S^a)$ is the nonlinear spin correction to the classical momentum equation,
 $\widetilde{\bm{\mathsf{\Sigma}}} = \langle(\bm{\nabla}\mathcal{S}_{(\alpha)a})\otimes(\bm{\nabla}\mathcal{S}^{a}_{(\alpha)}) \rangle$ is a pressure like spin term (which may be decomposed into trace-free part and trace),  
${\bm{\mathsf{K}}} = n_e\langle\bm{w}_{(\alpha)}\otimes\bm{\mathcal{S}}_{(\alpha)}\rangle$ is the 
thermal-spin coupling,  and $[(\bm{\nabla}\otimes\bm{B})\cdot\bm{S}\,]^a = (\partial^aB_b)S^b$.
Here the latin indices $a,b,\ldots = 1,2,3$ denote the vector components.
We note that the momentum conservation equation (\ref{eq:mom}) and 
the spin evolution equation (\ref{eq:spin-total}) still 
contains the explicit sum over the $N$ states, and (as in classical fluid theory) it is necessary 
to impose further
statistical relations in order to close the system \footnote{Using $L \gg \lambda_F$ where
$L$ is the typical fluid length scale and $\lambda_F$ is the Fermi wavelength, we obtain \cite{manfredi} $\langle\bm{\nabla}Q_{(\alpha)}\rangle \approx \bm{\nabla}\left[- 
    (\hbar^2/2m_en_e^{1/2})\nabla^2n_e^{1/2}\right]$. In the 
    Schr\"odinger treatment of quantum plasmas, this term is the only quantum contribution 
    to the equations of motion (see \cite{haas-etal1,anderson-etal,haas-etal2,haas, garcia-etal,marklund,manfredi}).
}. 
The preceding analysis applies equally well to electrons as holes or similar condensations. 
We will now include the ion species, which, due to the smaller charge-to-mass ratio, are described by the classical 
equations of motion.

The coupling between the quantum plasma species is mediated by the electromagnetic field. 
By definition we let $\bm{B}_{\mathrm{tot}}$ include spin sources, i.e.
$\bm{B}_{\mathrm{tot}} \equiv \bm{B} + \bm{B}_{\mathrm{sp}}$,
such that Ampere's law in terms of $\bm{B}_{\mathrm{tot}}$ reads 
$\nabla \times \bm{B}_{\rm tot} = \mu _{0}( \bm{j}+\bm{j}_{\mathrm{sp}}) 
  + c^{-2} \partial_t \bm{E}$, 
including the magnetization spin current 
$\bm{j}_{\mathrm{sp}}=\nabla\times(2n\mu_B \bm{S}/\hbar)$ 
\footnote{We note that this reshuffling of terms stems from the use of a 
  non-relativistic particle theory. Thus, it is expected that the need for such 
  rearrangements will disappear in a fully relativistic theory.}. 
We obtain consistency with the momentum conservation equation (\ref{eq:mom}) 
by adding a term proportional to $\bm{V\times B}_{\mathrm{sp}}$ to the
Lorentz force, and subtracting it from the quantum force. The above
alterations are only reshuffling of terms.
However, a difference do appear when closing the system
using Faraday's law. By letting
$\nabla \times \bm{E} = - \partial_t \bm{B}_{\mathrm{tot}}$,
using $\bm{B}_{\mathrm{tot}}$ instead of $\bm{B}$,  
we indeed obtain a difference compared to the classical Maxwell's equations. 
It is the full electromagnetic fields, including spin sources, that should be used in Faraday's law. Thus, Faraday's
law as presented here is therefore the correct one to use. This
form also gives a Hermitian susceptibility tensor (see below), something 
which is not obtained otherwise. 


To demonstrate the usefulness of the spin fluid equations, we 
investigate linear wave propagation in a magnetized plasma. For comparison 
we first neglect all quantum effects. Linearizing and Fourier analyzing the
equations of motion, and
substituting the velocities into Maxwells equations, we obtain 
$\bm{\varepsilon}\cdot \bm{E} = 0$, where 
\begin{equation}
\bm{\varepsilon} =  \bm{\mathsf{I}} 
  + \left( 
  \begin{array}{ccc}
  \chi_{\bm{\bot \bot }}  & \chi_{\bm{\bot }\top } & \chi_{\bm{\bot }z} \\ 
  -\chi_{\bm{\bot }\top } & \chi_{\top \top }         & \chi_{\top z} \\ 
  \chi_{\bm{\bot }z}       & -\chi_{\top z}            & \chi_{zz}
  \end{array}
  \right)   
  + \left( 
  \begin{array}{ccc}
  \frac{k_z^2c^2}{\omega^2} & 0                         & \frac{k_zk_{\perp}c^2}{\omega^2} \\ 
  0                            & \frac{k^2c^2}{\omega^2} & 0 \\ 
  \frac{k_zk_{\perp}c^2}{\omega^2}  & 0                                  & \frac{k_{\perp}^2c^2}{\omega^2}
  \end{array}
  \right) 
\label{dielectric tensor}
\end{equation}
and the standard susceptibility components are 
\begin{eqnarray}
&& \chi_{\bm{\bot \bot }} =-\sum_{\mathrm{p.s.}} \frac{\omega _{p}^{2}(\omega
^{2}-k_{z}^{2}v_{t}^{2})}{\omega _{\mathrm{w}}^{4}} , 
\nonumber \\ 
&& \chi_{\bm{\bot }\top } = -i\sum_{\mathrm{p.s.}} \frac{\omega
_{p}^{2}\omega _{c}(\omega ^{2}-k_{z}^{2}v_{t}^{2})}{\omega \omega _{\mathrm{%
w}}^{4}} , \quad 
\nonumber \\ 
&& \chi_{\bm{\bot }z} = -\sum_{\mathrm{p.s.}} \frac{\omega _{p}^{2}k_{\bm{\bot }}k_{z}v_{t}^{2}}{\omega _{%
\mathrm{w}}^{4}}, 
\nonumber \\ 
&& \chi_{\top \top } = - \sum_{\mathrm{p.s.}} \frac{\omega
_{p}^{2}(\omega ^{2}-k^{2}v_{t}^{2})}{\omega _{\mathrm{w}}^{4}} , \quad  \\  
&& \chi_{\top z} = i\sum_{\mathrm{p.s.}} \frac{\omega _{p}^{2}\omega _{c}k_{\bm{%
\bot }}k_{z}v_{t}^{2}}{\omega \omega _{\mathrm{w}}^{4}},\quad 
\nonumber \\ 
&& \chi_{zz} = - \sum_{\mathrm{p.s.}} \frac{\omega _{p}^{2}(\omega ^{2}-\omega _{c}^{2}-k_{\bm{%
\bot }}^{2}v_{t}^{2})}{\omega _{\mathrm{w}}^{4}}.
\nonumber
\end{eqnarray}
Here, the sums are over the particle species, $k=(k_{z}^{2}+k_{\bot }^{2})^{1/2}$, $\bm{k}_{\bot }$ is the
perpendicular (to $\widehat{\bm{z}}$) part of the wave-vector, the $\top 
$-direction is parallel to $\widehat{\bm{z}}\times \bm{k}_{\bot }$, $%
\omega _{p}$ is the plasma frequency ($\omega _{pe}$ for the electrons and $%
\omega _{pi}$ for the ions), $\omega _{c}=qB_{0}/m$ is the cyclotron
frequency, $q$ and $m$ are the particle charge and mass, $v_{t}^{2}$ is the
square of the thermal velocity times the ratio of specific heats, $%
c$ is the speed of light in vacuum, and $\omega _{\mathrm{w}}^{4}=\omega
^{2}\left( \omega ^{2}-k^{2}v_{t}^{2}\right) -\omega _{c}^{2}(\omega
^{2}-k_{z}^{2}v_{t}^{2})$. For notational convenience, the subscripts
denoting the various particle species have been left out. 

Next we determine the equilibrium
spin configuration. For many plasmas paramagnetic theory applies. Thus, 
in an external magnetic field $\bm{B}_{0} = B_0\widehat{\bm{z}}$, the zero order magnetization $%
\bm{M}_{S0}$ due to the spin can be written \footnote{The full magnetization 
  is known to include both Pauli spin paramagnetism and Landau orbit diamagnetism. 
  However, in Eq. (\ref{Spin-magn-relation}) only the Pauli contribution should be included}
\begin{equation}
  \bm{M}_{S0} = n_{0}\mu _{B}\, 
    \eta\!\left( \frac{\mu_{B}B_{0}}{KT}\right)\widehat{\bm{z}}  \label{Spin-magnetization}
\end{equation}
where $K$ is Boltzmann's constant, $T$ is the temperature, 
and we have introduced the Langevin function 
  $\eta(x) = [ \coth(x) - x^{-1}]$. 
Here we have assumed that the spin contribution to the total magnetic field
is small, otherwise $B_{0} \rightarrow B_{0} + B_{S0}$, where $B_{S0} =\mu
_{0}\mu _{B}n_{0}$, in Eq. (\ref{Spin-magnetization}). In general, the
spin-magnetization $\bm{M}_{S}$ and the spin-vector $\bm{S}$ are
related by 
\begin{equation}
  \bm{S}= \frac{\hbar\bm{M}_{S}}{2n\mu _{B}}
\label{Spin-magn-relation}
\end{equation}
and thus the zero order spin vector becomes 
$\bm{S}_{0} = (\hbar/2)\,\eta(\mu_{B}B_{0}/KT)\widehat{\bm{z}}$. 
From (\ref{Spin-magn-relation}) we obtain the spin-current contribution
$\bm{j}_{s}=\nabla \times ( 4n_ee\bm{S}/m_e)$.

Generalizing (\ref{dielectric tensor}) to include all terms from
quantum effects give extremely complicated expressions. However, for most
plasmas, the parameter $\mu _{B}B_{0}/KT$ is very small, the spins are
essentially randomly orientes, and the spin quantum effects are negligible.
On the other hand, for low-frequency wave motion in a highly magnetized (or
low temperature) plasma, the spin effects can be appreciable. In this case
the dominant contribution to the spin effects come from the component of
the spin force parallel to the magnetic field,   
$F_{Qz}=-({2\mu _{B}n_{0}S_{0}}/{\hbar })\partial _{z}B_{1}$, where $B_1$ denotes
the magnetic field perturbation, 
together with the part of the spin current in the $\top $-direction (from
the part proportional to $\nabla n\times \bm{S}_{0}$), and we can drop
all other components as well as quantum terms that are proportional to $%
\hbar ^{2}$, provided $eB_0 \gg \hbar k^2$. Keeping the above terms, including only the lowest order
contributions in $\omega /\omega _{ci}$, the susceptibility tensor is modified
to 
\begin{equation}
  \bm{\chi} = \left( 
  \begin{array}{ccc}
  \chi_{\bm{\bot \bot }} & \chi_{\bm{\bot }\top } & 
  \chi_{\bm{\bot }z} \\ 
  -\chi _{\bm{\bot }\top } & \chi_{\top \top } & 
  \chi_{\top z}+\chi_{\mathrm{sp}} \\ 
  \chi_{\bm{\bot }z} & -(\chi_{\top z}+\chi_{%
  \mathrm{sp}}) & \chi_{zz}
  \end{array}
  \right)
  \label{spin-dielectric tensor}
\end{equation}
where the spin-contribution 
is 
\begin{equation}
  \chi_{\mathrm{sp}}=i\,\eta\!\left( \frac{\mu_{B}B_{0}}{KT}\right) \frac{\omega _{p_{e}}^{2}\hbar k_{\bm{%
  \bot }}k_{z}}{\omega (\omega ^{2}-k_{z}^{2}v_{te}^{2})m_{e}}
  \label{spin-dielectric term}
\end{equation}
As an example we consider the fast and slow magnetosonic mode, which is now
described by the dispersion relation 
 $\varepsilon _{\top \top }\varepsilon _{zz}+(\varepsilon _{\top
z}+\varepsilon _{\mathrm{sp}})^{2}=0$. 
In the standard regime $\omega _{ci}^{2}/\omega _{pi}^{2}\ll 1$, $%
\omega \ll k_{z}v_{te}$, the dispersion relation becomes
\begin{equation}
(\omega ^{2}-k^{2}c_{A}^{2})(\omega ^{2}-k_{z}^{2}c_{s}^{2}) = \omega ^{2}k_{%
\bm{\bot }}^{2}c_{s}^{2}\left[ 1 + \eta\!\left( \frac{\mu_{B}B_{0}}{KT}\right) \frac{\hbar \omega _{ce}}{%
m_{e}v_{te}^{2}}\right]  \label{MS-DR}
\end{equation}
where the ion-acoustic velocity is $c_{s}=(m_{e}/m_{i})^{1/2}v_{te}$, the Alfv%
\'{e}n velocity is $c_{A}=(B_{0}^{2}/\mu _{0}n_{0}m_{i})^{1/2}$ and, for simpliticy,
we have assumed that the ion-temperature is smaller than the electron
temperature and included only electron thermal effects. Noting that $\mu
_{B}B_{0}/KT \equiv \hbar \omega _{ce}/m_{e}v_{te}^{2},$ obviously the spin-effects
are important if 
\begin{equation}
\frac{\hbar \omega _{ce}}{m_{e}v_{te}^{2}}\gtrsim 1 . \label{spin-condition}
\end{equation}
Thus, for laboratory magnetic fields, where at most $B_{0} \sim 10-20\,\mathrm{T}$,
we need low temperature plasmas for spin effects to influence the
fast and slow magnetosonic modes. However, in the vicinity of
pulsars and magnetars \cite{harding-lai}, we have $B_{0} \geq 10^{8}\,\mathrm{T}$. For such systems,
 spin plasma effects can be important even in a high temperature plasma.The spin effect 
on the fast and slow mode is illustrated in Fig.\ 1. 
Furthermore, we point out that for modes with even lower phase velocities (which
exist in for example dusty plasmas \cite{shukla-mamun}), the relative importance of the spin
susceptibility term is enhanced, and spin effects can be 
significant also under laboratory conditions. 

\begin{figure}
  \subfigure[]{\includegraphics[width=.67\columnwidth]{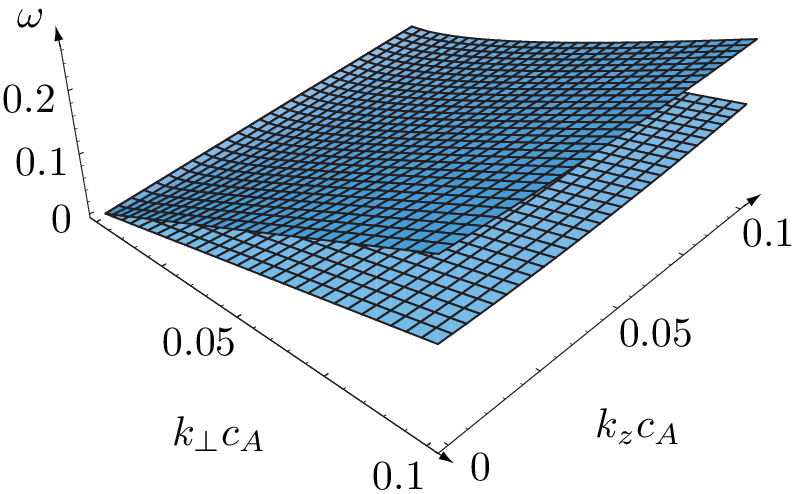}}\hspace{10mm}
  \subfigure[]{\includegraphics[width=.67\columnwidth]{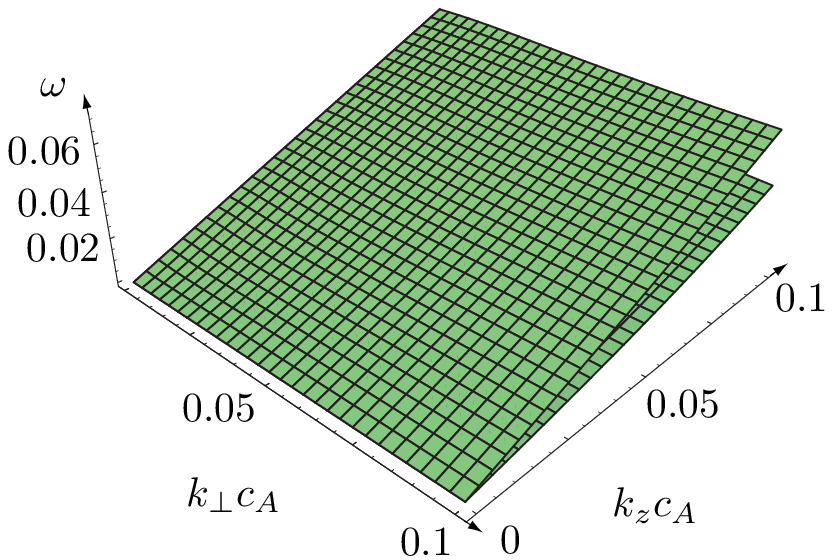}}
  \caption{The two roots of the dispersion relation (\ref{MS-DR}) plotted in (a) (fast mode) and (b) 
    (slow mode). In case (a) the lower surface is without spin and the upper surface is with spin,
    while in (b) the lower surface is with spin and the upper surface is without spin. 
    We note that the contribution from the spin term
    can be significant, in particular for large values of the wave numbers $k_z$ and
    $k_{\perp}$. Here we have used $c_s^2/c_A^2 = 0.5$, 
    $\eta\hbar \omega _{ce}/m_{e}v_{te}^{2} = 8$, and normalized 
    the frequency by the ion cyclotron frequency $\omega_{ci}$ and 
    the wave numbers by $\omega_{ci}/c_A$.}
\end{figure}

In conclusion, we have derived the multi-fluid equations for spin $\tfrac{1}{2}$ quantum plasmas, 
starting from the Pauli equation. In order to demonstrate the usefulness of our equations, we have analyzed the linear modes, and 
demonstrated that the low-frequency modes are significantly altered by the spin effects provided the condition (\ref{spin-condition})
is fulfilled. 
In many classical plasmas spin effects can be neglected due to the random orientations of the spin vector. We stress here, however, that our results show that the spin multi-fluid equations can have important applications to such different mediums as low-temperature solid state plasmas, as well as to the accretion discs surrounding pulsars and magnetars. 
Furthermore, we emphasize that the spin-contributions are typically more 
important than the usual quantum plasma corrections \cite{manfredi}, specifically when the inequality $eB_0 \gg \hbar k^2$ is fulfilled.        

The linearized results presented in this Letter will most likely find its experimental application in 
dusty plasmas, where the low phase velocity will make the relative importance of the spin contribution 
(\ref{spin-dielectric term}) particularly significant, enabling probing of the collective spin dynamics.

Finally, we suggest that the full nonlinear system (\ref{eq:density})--(\ref{omega}) will show interesting 
behavior close to the electrons cyclotron frequency, when the spin vector evolution becomes resonant.
Moreover, the importance of the pressure like spin terms for, e.g., astrophysical plasmas is a further field
of investigation.


\begin{thebibliography}{99}


\bibitem{haas-etal1}  F. Haas, G. Manfredi, and M. R. Feix, Phys. Rev. E 
\textbf{62}, 2763 (2000).

\bibitem{anderson-etal}  D. Anderson, B. Hall, M. Lisak, and M. Marklund,
Phys. Rev. E \textbf{65}, 046417 (2002).

\bibitem{haas-etal2}  F. Haas, L. G. Garcia, J. Goedert, and G. Manfredi,
Phys. Plasmas \textbf{10}, 3858 (2003).

\bibitem{haas}  F. Haas, Phys. Plasmas \textbf{12}, 062117 (2005).

\bibitem{garcia-etal}  L. G. Garcia, F. Haas, L. P. L. de Oliviera, and J.
Goedert, Phys. Plasmas \textbf{12}, 012302 (2005).

\bibitem{marklund}  M. Marklund, Phys. Plasmas \textbf{12}, 082110 (2005).

\bibitem{marklund-shukla}
M. Marklund and P. K. Shukla, Rev. Mod. Phys. \textbf{78}, 591 (2006).

\bibitem{Shukla-Eliasson}  P. K. Shukla and B. Eliasson, Phys. Rev. Lett. 
\textbf{96}, 245001 (2006).

\bibitem{Markowich-etal}  P. A. Markowich, C. A. Ringhofer, and C.
Schmeiser, \textit{Semiconductor equations} (Springer, Vienna, 1990).

\bibitem{Calvayrac-etal}  F. Calvayrac, P.-G. Reinhard, E. Suraud, and C.
Ullrich, Phys. Rep. \textbf{337}, 493 (2000).

\bibitem{Stenflo-etal}  L. Stenflo, P. K. Shukla, and M. Marklund, Europhys.
Lett. \textbf{74}, 844 (2006).

\bibitem{Becker-etal}  K. H. Becker, K. H. Schoenbach, and J. G. Eden, J.
Phys. D \textbf{39}, R55 (2006).

\bibitem{exp1}
Y. I. Salamin, S. X. Hu, K. Z. Hatsagortsyan, and C. H.
Keitel, Phys. Rep. 427, 41 (2006).

\bibitem{exp2} 
G. A. Mourou, T. Tajima, and S. V. Bulanov, \rmp \ \textbf{78}, 309 (2006).

\bibitem{shukla-stenflo}  P. K. Shukla and L. Stenflo, Phys. Lett. A, in
press (2006).

\bibitem{shukla-mamun}  P. K. Shukla and A. A. Mamun, \textit{Introduction
to Dusty Plasma Physics} (IOP Publishing, Bristol, 2002).

\bibitem{shukla}  P. K. Shukla, Phys. Lett. A \textbf{352}, 242 (2006).

\bibitem{manfredi}  G.\ Manfredi, quant-ph/0505004 (2005).

\bibitem{halperin-hohenberg}
B. I. Halperin and P. C. Hohenberg, Phys. Rev. \textbf{188}, 898 (1969).

\bibitem{blum}
F. A. Blum, Phys. Rev. B \textbf{3}, 2258 (1971).

\bibitem{balatsky}
A. V. Balatsky, Phys. Rev. B \textbf{42}, 8103 (1990).

\bibitem{rathe-etal}
U. W. Rathe, C. H. Keitel, M. Protopapas, and P. L. Knight, J. Phys. B: At. Mol. Opt. Phys.
\textbf{30}, L531 (1997).

\bibitem{hu-keitel}
S. X. Hu and C. H. Keitel, Phys. Rev. Lett. \textbf{83}, 4709 (1999).

\bibitem{arvieu-etal}
R. Arvieu, P. Rozmej, and M. Turek, Phys. Rev. A \textbf{62}, 022514 (2000).

\bibitem{aldana-roso}
J. R. V\'azquez de Aldana and L. Roso, J. Phys. B: At. Mol. Opt. Phys.
\textbf{33}, 3701 (2000).

\bibitem{walser-keitel}
M. W. Walser and C. H. Keitel, J. Phys. B: At. Mol. Opt. Phys. \textbf{33}, L221 (2000).

\bibitem{walser-etal}
M. W. Walser, D. J. Urbach, K. Z. Hatsagortsyan, S. X. Hu, and C. H. Keitel,
Phys. Rev. A \textbf{65}, 043410 (2002).

\bibitem{qian-vignale}
Z. Qian and G. Vignale, Phys. Rev. Lett. \textbf{88}, 056404 (2002).

\bibitem{roman-etal}
J. S. Roman, L. Roso, and L. Plaja, J. Phys. B: At. Mol. Opt. Phys. \textbf{37}, 435 (2004).

\bibitem{liboff}
R. L. Liboff, Europhys. Lett. \textbf{68}, 577 (2004).

\bibitem{fuchs-etal}
J. N. Fuchs, D. M. Gangardt, T. Keilman, and G. V. Shlyapnikov,
Phys. Rev. Lett. \textbf{95}, 150402 (2005).

\bibitem{melrose}
D. B. Melrose and A. J. Parle, Aust. J. Phys. \textbf{36}, 755 (1983);
D. B. Melrose, \textit{ibid.}, 775 (1983);
D. B. Melrose and A. J. Parle, \textit{ibid.}, 799 (1983).

\bibitem{harding-lai}
A. K. Harding and D. Lai, Rep. Prog. Phys. \textbf{69}, 2631 (2006).

\bibitem{melrose-weise}
D. B. Melrose and J. I. Weise, Phys. Plasmas \textbf{9}, 4473 (2002).

\bibitem{baring-etal}
M. G. Baring, P. L. Gonthier, and A. K. Harding, Astrophys. J. \textbf{630}, 430 (2005).

\bibitem{brodin-etal}
G. Brodin, M. Marklund, L. Stenflo, and P. K. Shukla, New J. Phys. \textbf{8}, 16 (2006).

\bibitem{holland}
P. R. Holland, \textit{The Quantum Theory of Motion} (Cambridge University Press, Cambridge, 1993).

\end{thebibliography}
\end{document}